\documentclass[runningheads]{llncs}
\usepackage[T1]{fontenc}

\usepackage{float} 
\usepackage{graphicx}
\usepackage{subcaption}
\usepackage[hidelinks]{hyperref}
\usepackage{color}
\usepackage{orcidlink}

\usepackage{listings}
\usepackage{wrapfig}

\begin{document}

\title{Accelerating Cloud-Based Transcriptomics: Performance Analysis and Optimization of the STAR Aligner Workflow}
\titlerunning{Accelerating Cloud-Based Transcriptomics}

\author{Piotr Kica\inst{1,2}\orcidlink{0009-0001-9269-6804} \and
Sabina Lichołai\inst{1,3}\orcidlink{0000-0001-7008-2492} \and \\
Michał Orzechowski\inst{1,2,3}\orcidlink{0000-0002-8558-1283} \and 
Maciej Malawski\inst{1,2}\orcidlink{0000-0001-6005-0243} \\
\texttt{\{p.kica, s.licholai, \\ m.orzechowski, m.malawski\}@sanoscience.org}
}

\authorrunning{P. Kica et al.}

\institute{Sano Centre for Computational Medicine, Krak\'ow, Poland \and
Faculty of Computer Science, AGH University of Krak\'ow, Poland \and
Academic Computer Centre Cyfronet AGH, Krak\'ow, Poland}

\maketitle

\begin{abstract}
In this work, we explore the Transcriptomics Atlas pipeline adapted for cost-efficient and high-throughput computing in the cloud. We propose a scalable, cloud-native architecture designed for running a resource-intensive aligner -- STAR -- and processing tens or hundreds of terabytes of RNA-sequencing data. We implement multiple optimization techniques that give significant execution time and cost reduction. The impact of particular optimizations is measured in medium-scale experiments followed by a large-scale experiment that leverages all of them and validates the current design. Early stopping optimization allows a reduction in total alignment time by 23\%. We analyze the scalability and efficiency of one of the most widely used sequence aligners. For the cloud environment, we identify one of the most suitable EC2 instance types and verify the applicability of spot instances usage.

\end{abstract}

\section{Introduction}
\label{sec:introduction}

The cloud is, in many cases, the infrastructure of choice for large-scale genomic pipelines, as it promises scalability, cost efficiency, and availability of on-demand resources. There are multiple recent examples of pipelines that have been developed for genomic data, built directly using cloud services ~\cite{wiewiorka2023cloud,camacho2023elasticblast}. Such cloud-native pipelines take advantage of the capabilities of cloud services, allowing parallelism, scalability, and elasticity, often with autoscaling, and follow the principles of infrastructure as a code for application deployment and setup of the environment. 

However, cloud infrastructure is complex, as there are multiple services offered with a wide range of configuration options, so exploiting the full potential of clouds requires combining application-specific expertise with the ability to fine-tune cloud infrastructure configuration. For example, there are multiple compute service options (IaaS, CaaS, FaaS), storage systems (Object Storage, NFS filesystems, EBS-like volumes), database, and queuing/messaging systems, etc., so making the right choice becomes a challenge.

In this paper, we present a case study of running a Transcriptomics Atlas pipeline in the AWS cloud. It is a data- and compute-intensive pipeline, based on a sequence aligner -- STAR~\cite{dobin2013star} -- that processes hundreds of terabytes of RNA-seq data. We show that designing a cost-efficient system is nontrivial and requires thoughtful decisions and optimizations. Specifically, our aim is to answer relevant research questions, which can be grouped into application-specific and infrastructure-specific ones:\\

\noindent 
\textbf{Application-specific research questions:}
\begin{itemize}
    \item How to take advantage of the intermediate results to reduce time and cost?
    \item How to select the optimal level of parallelism within a single node?
    \item How to distribute the initial input data (STAR Index) to compute instances?
\end{itemize}
\textbf{Cloud infrastructure research questions:}
\begin{itemize}
    \item Which instance types are the most cost-efficient for alignment?
    \item How suitable are spot instances for running resource-intensive aligners?
\end{itemize}

Our research advances the field by analyzing the \textit{early stopping} feature which significantly increases the throughput of the alignment. We evaluate the scalability of STAR in order to find the most cost-efficient allocation of cores. In our solution, we encounter and solve the problem of STAR index distribution to worker instances. Cloud-related optimizations, such as finding the most suitable instance type for STAR alignment and spot instance usage applicability, allow for a decrease in the cost of running the pipeline. We focus on STAR alignment workflow in the cloud; however, many of our findings can be applied in other environments (HPC) and be useful for optimizing other transcriptomics workflows.

The paper is organized as follows. In Section~\ref{sec:background} we provide the background information about the Transcriptomics Atlas and the software tools we use. Next, in Section~\ref{sec:related} we discuss the related work regarding the processing of RNA sequencing data in the cloud, and cloud architectures in general. Section~\ref{sec:architecture} presents the cloud-native architecture of the pipeline based on AWS cloud services. Based on that, we introduce application-specific optimizations in Section~\ref{sec:application-specific optimizations}, and cloud infrastructure optimizations in Section~\ref{sec:cloud infrastructure optimizations}. All optimizations are included in a large-scale experiment presented in Section~\ref{sec:large-scale experiment}. We provide the conclusions in Section~\ref{sec:conclusions} and outline future work in Section~\ref{sec:future}.

\section{Background}
\label{sec:background}

\subsection{Transcriptome and NGS sequencing}
The human transcriptome, which includes the complete set of RNA molecules present in human cells, constitutes a significant area of inquiry within the realms of molecular biology and medicine~\cite{pertea2012human}. It consists of different types of RNA, ranging from messenger RNA (mRNA), which is responsible for conveying genetic information from DNA to proteins, to various forms of non-coding RNA that play pivotal regulatory roles in biological processes~\cite{gibbs2020human}.

Exploring the human transcriptome is an important aspect of gaining deeper insight into the complexity of biological processes that occur in the organism and can help identify potential pathogenic factors and therapeutic targets. The transcriptome reflects both the natural genetic diversity of populations and the variability arising from various physiological and pathological states ~\cite{porcu2021differentially}.

Analyses utilizing transcriptomics are typically conducted in a comparative context, where different health states, diseases, or stimuli are examined in comparison to baseline conditions. However, the need to obtain data under standard conditions significantly increases the cost of experiments~\cite{casamassimi2017transcriptome}. Furthermore, the lack of such data during experiment planning poses challenges in optimizing experimental conditions. In response to these challenges, we took on the task of designing the Transcriptomics Atlas, where data from a representative collection of human tissues were processed in a uniform manner. These data enable broad applications in scientific research, including pharmacogenomics and biomarker research.

Currently, the most widely used technology for obtaining the human transcriptome is second-generation sequencing, based on the generation of short reads of RNA sequences. This technology enables rapid and precise identification of the transcriptome composition within a biological sample. However, due to the limited length of reads, subsequent assembly into a complete transcriptome necessitates specialized bioinformatics algorithms in a process called alignment.

The alignment process plays a pivotal role in the analysis of transcriptomic data acquired through sequencing technologies. A reference genome serves as the foundational scaffold for the alignment process, providing a comprehensive representation of the genetic material of a species. It acts as a framework against which experimental data can be compared and interpreted~\cite{satam2023next}. The alignment process entails computationally aligning short reads to the reference genome, to identify their precise genomic locations and infer transcriptional events~\cite{alser2021technology}. 
To facilitate alignment, various resources are available, among which the Ensembl database stands out as a significant repository of reference genomes and related genomic data. 

\subsection{Software used in the pipeline}
STAR~\cite{dobin2013star,STAR_github} is a software that allows the alignment of large ($>$80 billion reads) transcriptome RNA-sequences. It is a well-established and accurate aligner~\cite{baruzzo2017simulation,bianchi2023comparing} in bioinformatics written in C/C++. Usually, it requires a large amount of RAM to run (tens of GiBs, depending on the size of a reference genome) and a high-throughput disk to scale efficiently with an increasing number of threads. This aligner, as many others, requires a precomputed genomic \emph{index} data structure. 

The SRA-Toolkit~\cite{SRA-Toolkit} is a collection of tools that allow the access and handling of RNA-seq files stored in the NCBI SRA database~\cite{NCBI}. Among these tools, prefetch retrieves raw \textit{SRA} files from the NCBI database. Fasterq-dump converts \textit{SRA} files into \textit{FASTQ} format, which is a format accepted by STAR for alignment.

\section{Related work}
\label{sec:related}

\subsection{RNA-sequencing in the cloud}
Many efforts have been put into efficient and scalable RNA sequence processing. In~\cite{zou2021parallel}, authors explore different parallel computing strategies and limitations for multiple genomic tools. They identified usefulness in architecture-aware and also data-storage optimizations. Research on STAR-based RNA-seq processing workflow as well as cost and throughput analysis for cloud and HPC experiments, are carried out in~\cite{wilks2021recount3}. Pseudoaligners such as Salmon~\cite{salmon_paper} and Kallisto~\cite{bray2016near} are recommended by~\cite{lachmann2020interoperable} when cost plays a critical role. The authors were also able to create a cost-efficient, microservices-based architecture in the cloud for a pseudo-alignment workflow. Research carried out in~\cite{cinaglia2023massive} shows that serverless computing for RNA sequencing is a valid approach when high parallelism is the end goal. They were able to create a split-align-merge solution with HiSat2 running on multiple AWS Lambda instances and acquire 79\%-96\% time reduction compared to a local environment. Similarly, in~\cite{hung2020accessible}, authors used AWS Lambda with Burrows-Wheeler Aligner and processed their dataset in 1 minute instead of 19 hours. However, deploying STAR to serverless services may be more challenging compared to less resource-intensive aligners. Furthermore, in~\cite{arjona2023scaling} authors moved an HPC workflow to serverless services and identified multiple challenges with efficient data partitioning and transfer, insufficient object storage performance, and statelessness of functions.

Our previous work regarding the Transcriptomics Atlas has been included in~\cite{bader2023novel}. Authors compared an earlier version of the Transcriptomics Atlas pipeline architecture in both cloud and HPC environments. Their work focused on understanding the pipeline requirements for the pseudoalignment path which has common steps with the full alignment path. They measured resource usage for every step of the pipeline, tested the architecture in both cases, and compared performance between cloud and HPC. 

\subsection{Cloud batch systems}
Cloud batch processing leverages the scalability, flexibility, and resources of cloud computing platforms to process large volumes of data or perform computationally intensive tasks. Major public cloud providers offer dedicated batch systems, such as AWS Batch or Microsoft Azure Batch, which are similar alternatives to the proposed solution and could be used to deploy the Transcriptomics Atlas pipeline and process the input dataset.
Private clouds can leverage open source solutions like Apache Spark and Apache Airflow, or when using Kubernetes as a resource manager, native Kubernetes solutions such as Argo Workflows or KubeFlow, which has been tested in bioinformatics scenarios~\cite{yuan2020bioinformatics}.

\section{Pipeline and cloud architecture}
\label{sec:architecture}

\subsection{Pipeline description}
\begin{figure*}[h]
    \centering
    \includegraphics[scale=0.48]{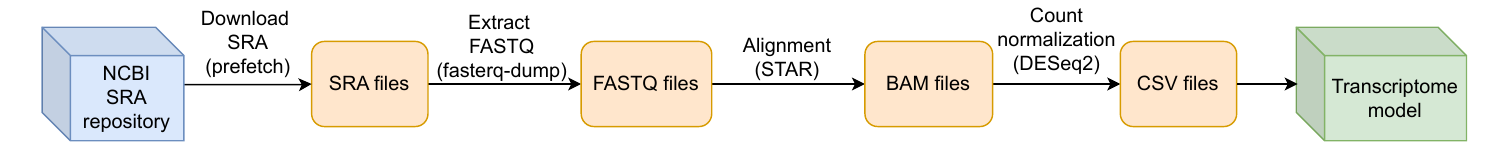}
    \caption{Transcriptomics Atlas Pipeline.}
    \label{fig:TAtlas_pipeline}
\end{figure*}

The general pipeline is presented in Fig.~\ref{fig:TAtlas_pipeline}. The first phase consists of accessing an \textit{SRA} file using \texttt{prefetch} and converting it into \textit{FASTQ} with \texttt{fasterq-dump}. The most important and time-consuming step is the alignment with STAR. Finally, the acquired \textit{BAM} file is normalized using \texttt{DESeq2} and stored. Instead of STAR, one can use alternative aligners (such as HiSat2) or pseudoaligners (such as Salmon). We are running STAR version 2.7.10b with \mbox{"\texttt{--quantMode~GeneCounts}"} option. STAR aligner gives us highly reliable results and allows extensive customization of alignment parameters. The pipeline steps are connected using a Python script, and the current implementation of the Transcriptomics Atlas project is publicly available on GitHub under the MIT license~\cite{neardata_repo}. 

\subsection{Input dataset}
\label{sec:input_dataset}

\begin{figure}[H]
    \centering
    \includegraphics[scale=0.39]{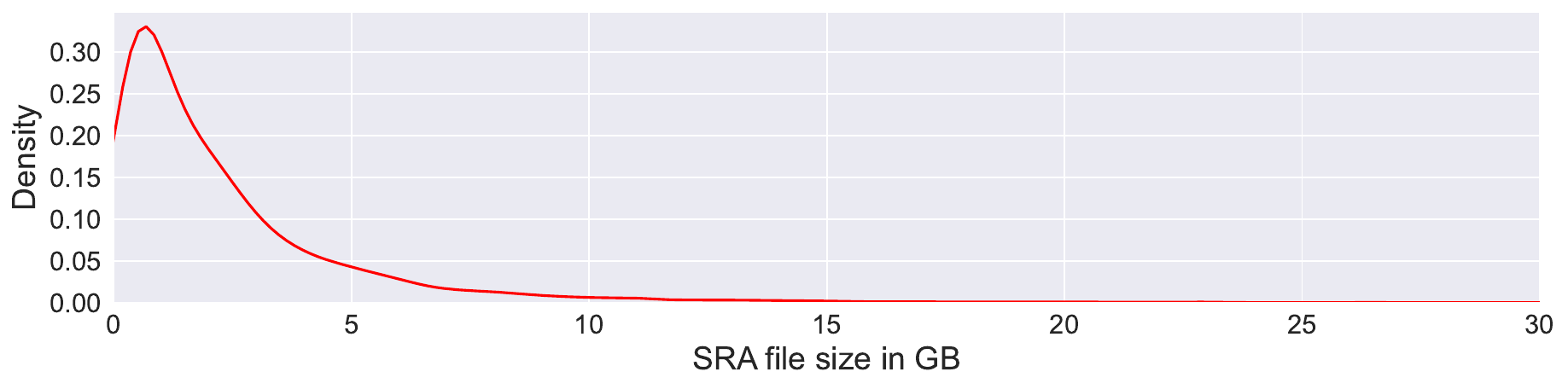}
    \caption{Distribution of input dataset.}
    \label{fig:input_dataset_density}
\end{figure}

Data are obtained from the NCBI Sequence Reads Archive repository~\cite{NCBI_SRA}, which contains more than 30PB of sequencing data. We select nucleotide sequence data generated by human samples on the basis of tissue type and appropriate technical parameters. The files themselves are hosted, i.a., on AWS (in the us-east-1 region). The input dataset is collected by querying the NCBI SRA database. We specify that the query should return only publicly available data of human origin and sequenced using Illumina machines. However, we noticed that the query can return false positives (e.g. data of non-human origin), therefore, additional filtering steps may be required. After querying the NCBI database, we download all available metadata for available SRA IDs for a given tissue and select IDs with compressed sequence size in the 200MB - 30GB range. We define the range by taking into account the size of libraries for typical sequencing in the context of the transcriptome, as well as output from the most commonly used wet lab sequencing protocols. This selected 218 TB (approximately 100.000 files) of relevant \textit{SRA} data for about 19 different tissues. For Transcriptomics Atlas, we aim to acquire 100-200 files (with a good mapping rate) per tissue. By selecting up to 400 samples for each tissue, the resulting input dataset consists of about 7250 files, which is 17TB of \textit{SRA} data. The size distribution of this subset is presented in Fig.~\ref{fig:input_dataset_density}. About 97\% of all input files are smaller than 10GB.

\vspace{-1cm}
\subsection{Resource requirements}
\label{sec:resource_requirements_section}

STAR is a resource-intensive aligner that uses a precomputed genome index in the alignment step that has to be loaded into the system memory. The generation of such an index is a one-time task. Depending on the type and release version of the genome, the index differs in size. 
We use the human genome of the "Toplevel" type distributed by Ensembl~\cite{ensembl_2023}. Previous work~\cite{kica2024optimizing} showed that an older release (108) results in an index of 85GiB in size, and using a newer one (release 111) is much smaller (29.5GiB) and faster (12 times). 
Also, STAR requires additional memory for sorting a BAM file, usually 1-2 GiB, but outliers may require even 20.5 GiB.

The pipeline requires enough disk space to handle intermediate files such as \textit{FASTQ} files along with \textit{SRA} and \textit{BAM} files. 
The fasterq-dump tool creates \textit{FASTQ} files, on average, 7.5 times bigger than the original \textit{SRA} file, however outliers can be even 17 times larger. Moreover, additional space is required during conversion.
This use case focuses on files within the 200MB - 30GB range. Therefore, we can estimate that the required space should not exceed 550GiB. Also, the disk should give sufficient performance (throughput, IOPS) to run the pipeline efficiently. We focus on the cost-efficiency of processing a large number of files, and processing time is of lower importance.

\subsection{Cloud architecture}
\begin{figure*}[h]
    \centering
    \includegraphics[scale=0.62]{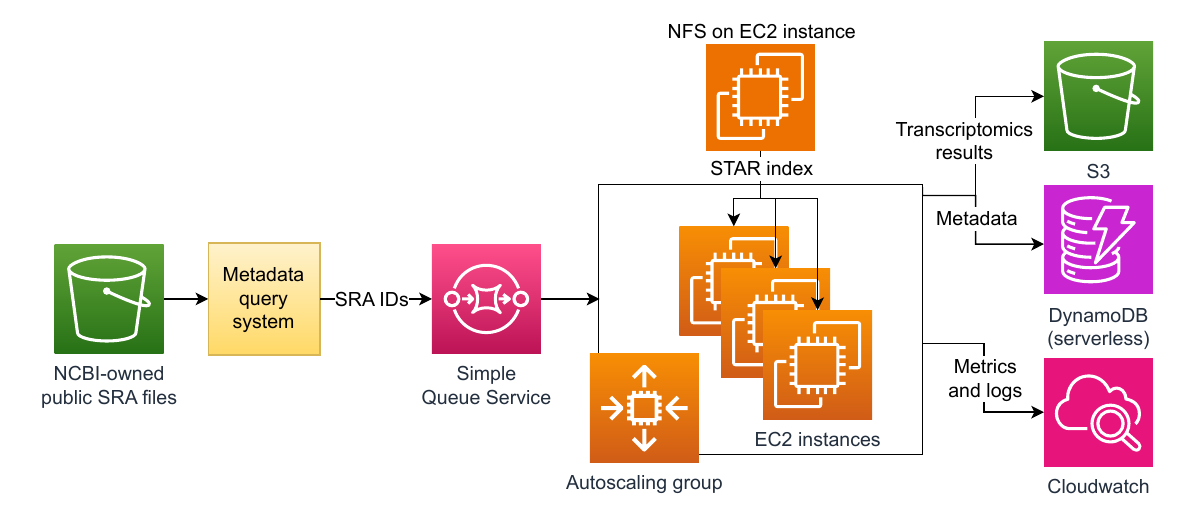}
    \caption{Cloud architecture for Transcriptomics Atlas Pipeline.}
    \label{fig:cloud_architecture}
\end{figure*}

Cloud architecture for the Transcriptomics Atlas pipeline is presented in Fig.~\ref{fig:cloud_architecture}. The main processing is done on a virtual cluster of EC2 spot instances (workers) launched from a custom machine image containing all the required software. The initial step for each worker is to connect to an NFS instance and load the STAR index into memory. Once initialization has been completed, workers acquire the SRA IDs from the queue and process them using the pipeline presented in Fig.~\ref{fig:TAtlas_pipeline}. The transcriptomics results are stored in a dedicated S3 bucket. Execution metadata are gathered for performance analysis and saved in a Dynamodb table. The table is used to prevent the processing of files that have already been processed. Metrics are saved using the CloudWatch service which gives live insight into resource utilization. The deployment of the infrastructure is automated with Terraform. 

The proposed approach is easily scalable and adaptable for similar workflows. Having extensive control over the underlying compute resources allows us to fine-tune the configuration for given requirements, which improves cost-efficiency. 

\section{Application-specific Optimizations}
\label{sec:application-specific optimizations}

\subsection{Early stopping during alignment}
Early stopping is a common method to stop the training of machine learning models when desired accuracy is achieved. We apply a similar approach to alignment by discarding low-quality or invalid sequences during processing. One of the features that makes STAR suitable for high-throughput computing is the reporting of the intermediate mapping rate, which allows the identification of such sequences. As presented in~\cite{kica2024optimizing}, utilization of live metrics from \textit{Log.progress.out} file can lead to a significant increase in throughput in the alignment phase (even 19.5\%). This feature is beneficial when we cannot (with an acceptable margin) determine the quality of the FASTQ beforehand, as is the case with our data set.

For the Transcriptomics Atlas project, the mapping rate threshold is set at 30\%. However, this is highly dependent on the use case. Pipelines that utilize STAR in a similar scenario and require sequences of the highest quality will greatly benefit from this feature. 
This encompasses standard scenarios involving the analysis of gene expression alterations due to various environmental factors, exogenous xenobiotics, or other stimulants. In addition, it will be applicable to studies involving tissues obtained from diseased organs. However, it should be noted that this threshold can be modified depending on the specific use case. For example, in studies aimed at the identification of novel RNA molecules or isoforms of existing ones, the threshold should be set lower, allowing for greater flexibility. In contrast, for immunological studies and analysis of data from specific libraries such as TCRs, the requirement for accurate annotation is increased, and the threshold should be set more stringently. 

\begin{wrapfigure}[17]{R}{0.57\textwidth}
    \centering
    \vspace{-0.7cm}
    \hspace{-0.7cm}
    \includegraphics[scale=0.5]{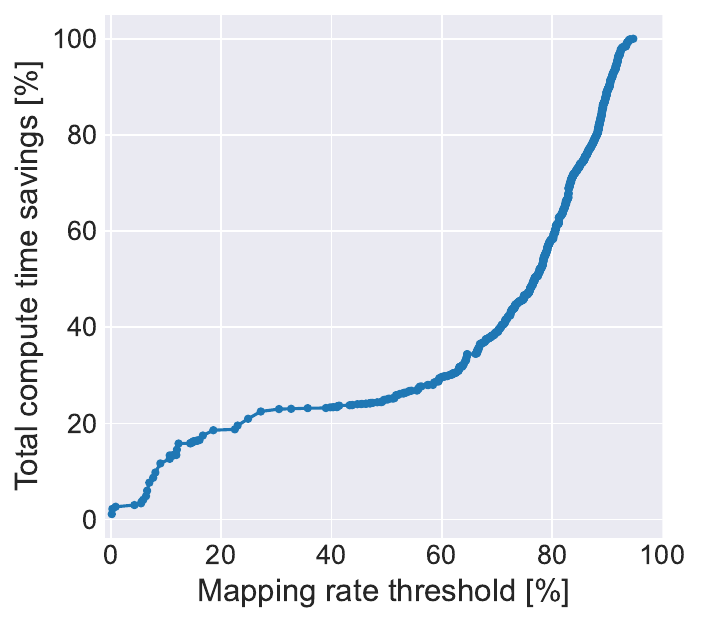}
    \caption{Impact of early stopping threshold on total alignment time.}
    \label{fig:mr_threshold}
\end{wrapfigure}

In Fig.~\ref{fig:mr_threshold}, we present an extended analysis based on the experiment carried out in~\cite{kica2024optimizing}. If we set the threshold at 80\%, we would reduce the total compute time by 60\%. The minimal number of processed spots was set to 10\%, but can be adjusted according to needs. As stated in~\cite{kica2024optimizing}, the inputs identified for termination turned out to be large single-cell sequences. However, it is possible to exclude correctly described single-cell sequences on the basis of metadata for a given SRA ID. Future work will include both approaches.

\subsection{Cost-efficient allocation of cores}
\label{sec:allocation_of_cores}

The recommended approach for STAR aligner is to match the thread count with the number of available cores in the node. Also, authors of the original STAR publication \cite{dobin2013star} claim that "STAR exhibits close to linear scaling of the throughput rate with the number of threads, losing about 10\% of per thread mapping speed when the number of threads is increased from 6 to 12". However, the original research lacks detailed performance and scalability analysis, which is important to maximize CPU efficiency. We decided to test the scalability of STAR to confirm these claims and find an optimal number of cores per worker, which will directly improve the throughput of the pipeline. The test suite consists of 3 input \textit{FASTQ} files of different sizes. Execution times are measured on two different 16-vCPU instance types - with and without Simultaneous Multi-Threading (SMT). The instances use AMD EPYC processors as these types turned out to be the most cost-efficient in the~\ref{sec:instance_type} section. The attached block storage with 1000 MiB/s throughput and 4000 IOPS is more than sufficient for this test.

\begin{figure}[H]
    \centering
    \includegraphics[width=\textwidth]{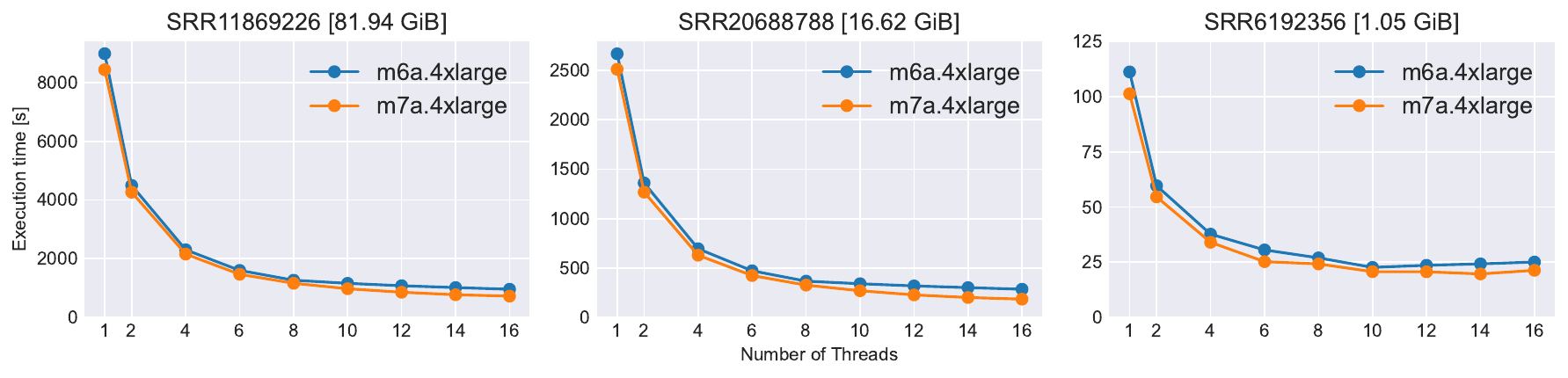}
    \caption{Scalability of the STAR aligner.}
    \label{fig:STAR_Scalability}
\end{figure}

\begin{figure}[H]
    \centering
    \includegraphics[width=\textwidth]{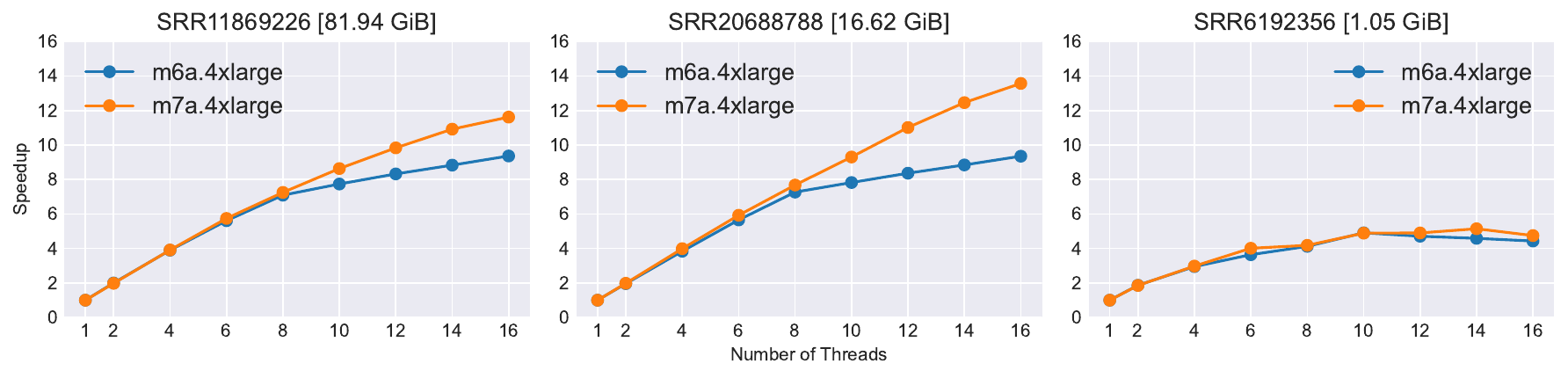}
    \caption{Speedup of the STAR aligner.}
    \label{fig:STAR_Speedup}
\end{figure}

\begin{figure}[H]
    \centering
    \includegraphics[width=\textwidth]{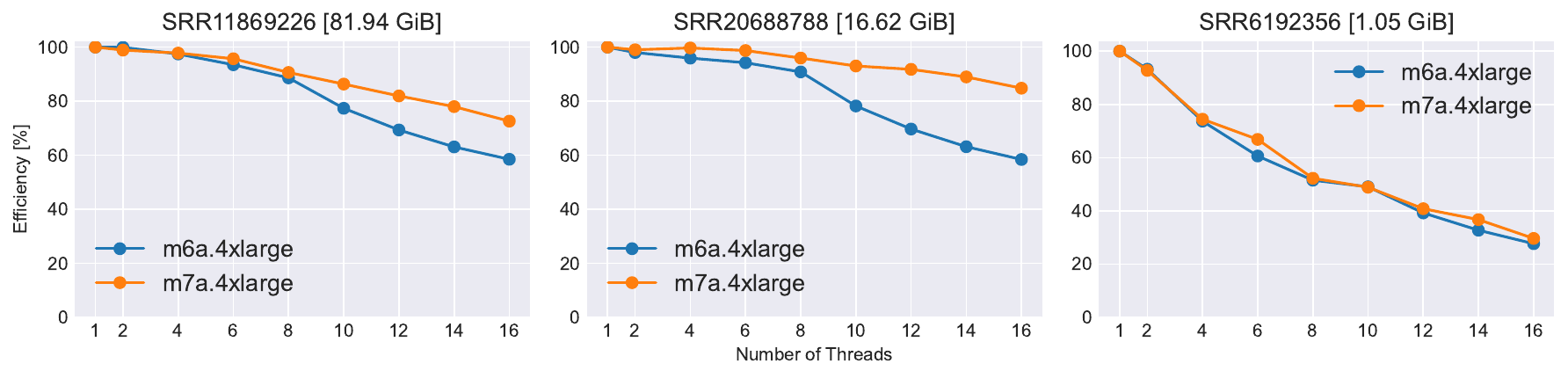}
    \caption{Efficiency of the STAR aligner.}
    \label{fig:STAR_Efficiency}
\end{figure}

The test results are presented in Figs.~\ref{fig:STAR_Scalability},~\ref{fig:STAR_Speedup},~and~\ref{fig:STAR_Efficiency}, and we see the benefit of the increased number of threads. However, there is a noticeable drop in efficiency - for 16 threads and m7a.large instance, we get 84\% and 72\% efficiency for 2GiB file and 10GiB respectively. This is especially visible for the m6a.large instance, which uses SMT and exceeding 8 threads further decreases efficiency. Based on the acquired metrics, we decided to focus on 8-vCPU instances for the best cost-efficiency.
We extend the study of A. Dobin et al.~\cite{dobin2013star}, by analyzing the performance on different file sizes, provide a detailed impact of the increase in core allocation and the effects of using SMT. Insights from this research are also useful for other environments (e.g., HPC centers).

\vspace{-0.9cm}

\vspace{-0.9cm}

\subsection{STAR index distribution}
In order to create a cost-efficient and high-throughput computing architecture it is necessary to find a viable solution for the distribution of the pre-computed STAR index to each worker. The goal is to be able to spin up tens of instances instantly and after processing the given workload, turn them off. The implemented approach utilizes an NFS server in the same region where workers are launched. A single c6in.xlarge is sufficient to distribute the 29.5GiB index to 50 workers in 14 minutes. This solution is easy to launch, relatively portable, and scalable (larger instance with higher bandwidth). The big advantage of mounting a file system is that each worker can simply load the index directly to memory without copying it to local storage. Running this instance for the duration of the whole experiment is a negligible part of the costs. The resources are heavily utilized at the start and underutilized most of the remaining time. However, we need to pay for the data transfer across the availability zones in the cloud. In the case of AWS it is currently priced at 0.01~\$ per GB~\cite{AWS_data_transfer_docs}, which results in about \$0.30 worker initialization cost. This solution is viable, vendor-independent, and acceptable in smaller-scale scenarios. However, further work is required to minimize this cost on a larger scale. 

\section{Cloud Infrastructure Optimizations}
\label{sec:cloud infrastructure optimizations}

\subsection{Instance type comparison in AWS}
\label{sec:instance_type}

According to the requirements in Section~\ref{sec:resource_requirements_section} and knowing that STAR is a memory intensive program, we decided to focus on instances with a higher memory per CPU factor with at least 64GiB of RAM. Using instance types which have more cores but less memory per CPU may require faster block storage and result in increased under-utilization during other, much less CPU-intensive steps (e.g. \textit{prefetch}). Using more cores in a single worker node would also reduce CPU efficiency during alignment as described in Section~\ref{sec:allocation_of_cores}.

The selected instance types of the current generation that meet these requirements are compared in Table~\ref{tab:instance_types}. 
This table also presents the total cost and time for performing STAR alignment on 50 random \textit{FASTQ} files. Only STAR processing time is considered - other pipeline steps account for only 24-31\% of total execution time and are less affected by additional/faster resources. The results indicate r7a.2xlarge as the fastest and cheapest type. However, when using spot instances, one should also take into consideration the spot availability of a given type.

\begin{table}[h]
\tiny
\sffamily
\caption{Cost-efficiency analysis of selected instance types.}
\resizebox{\textwidth}{!}{%
    \centering

    \begin{tabular}{|c|c|c|c|c|c|l|}
    \hline
    \begin{tabular}[c]{@{}c@{}}Instance\\  type\end{tabular} & vCPU & Cores & RAM [GiB] & \begin{tabular}[c]{@{}c@{}}On-demand \\ price [h]\end{tabular} & \begin{tabular}[c]{@{}c@{}}Total STAR \\ execution time [h]\end{tabular} & \begin{tabular}[c]{@{}l@{}}Total\\  cost\end{tabular} \\ \hline
    r6a.2xlarge   & 8    & 4     & 64  & \$0.4536            & 8.00                                                             & 3.63 \$    \\ \hline
    r6i.2xlarge   & 8    & 4     & 64  & \$0.5040            & 8.04                                                             & 4.05 \$    \\ \hline
    r7a.2xlarge   & 8    & 8     & 64  & \$0.6086            & 5.48                                                              & 3.33 \$    \\ \hline
    r7i.2xlarge   & 8    & 4     & 64  & \$0.5292            & 7.66                                                             & 4.05 \$    \\ \hline
    \end{tabular}
    \label{tab:instance_types}
}
\end{table} 

\pagebreak

\subsection{Cost-efficiency of spot instances}
\label{sec:cost_efficiency_of_spot_instances}

Spot instances on AWS offer compute resources at a lower cost (depending on instance type and current market demand). However, such instances can be terminated with a 2 minute notice. For example, an r7a.2xlarge instance can be acquired with 50\%-60\% discount. The availability for a given type and discounted price change over time. 

Our use case fits this model because with previously described optimizations, the pipeline runs relatively quickly (mean = 8 min). Currently, there is no good way to efficiently checkpoint the computations, so if the termination happens, the file must be processed once again from the start. Moreover, each interruption results in requesting and creating a new instance; therefore, the STAR index must be once more loaded to the memory of the worker. This means that the high interruption rate is even less desirable. In our use case, it is more cost efficient to use instances with a lower interruption rate than with a higher discount.

\begin{figure}[H]
    \centering
    \includegraphics[scale=0.49]{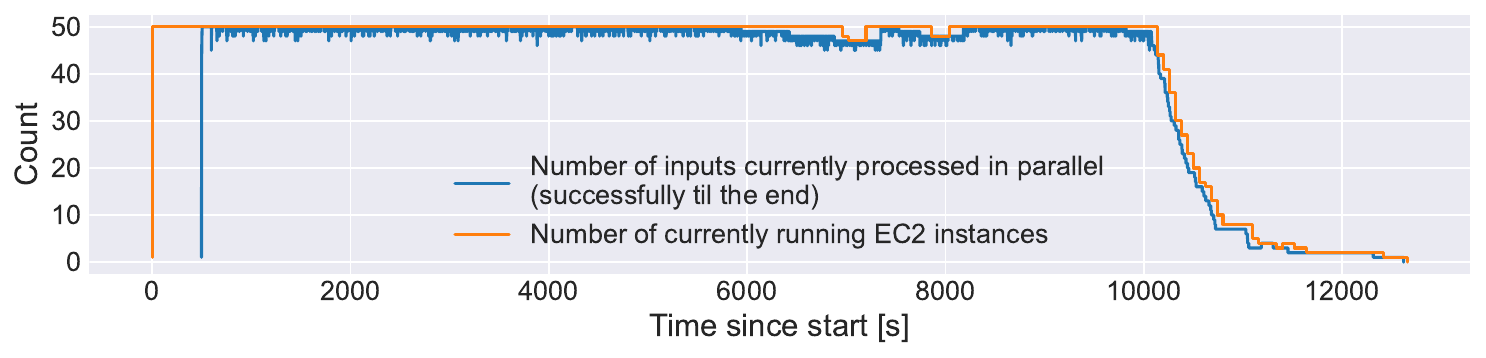}
    \caption{Spot instances usage experiment. Timeline for instances and successfully computed files. Visible initialization phase and 5 interruptions.}
    \label{fig:optimistic_spot_ec2_timeline}
\end{figure}

However, with good configuration (for example, using instance type with low interruption rate), using spot instances should result in relatively stable computations, as shown in Fig.~\ref{fig:optimistic_spot_ec2_timeline}. Here, we present the results of processing 1000 \textit{SRA} files on r7a.2xlarge instances. During the experiment, only five interruptions occurred and resulted in a loss of less than 1\% of the total running time among all instances. It also shows the initialization phase, where it takes 8.3 min for 50 instances to download and load the 29.5GiB index to memory. 

\section{Large-scale Experiment}
\label{sec:large-scale experiment}

\subsection{Goal and configuration}
The goal of the experiment is to test the pipeline on a larger scale, measure, and analyze resource usage. We use an input data set described in Section~\ref{sec:input_dataset}. We expect the results to show the impact of the optimizations implemented.
The configuration of the experiment is listed below.
\begin{itemize}
    \small
    \item EC2 (Spot): 50 r7a.2xlarge instances in us-east-1
    \item EBS: 550GB, GP3, 500MiB/s, 3000IOPS
    \item Input: 7216 \textit{SRA} files (2.5GB avg, 17.9TB total size, max=29.9GB)
    \item Index: Based on Toplevel human genome, release 111, 29.5GB size.
\end{itemize}

\subsection{Results}

The experiment has been completed successfully giving new insight into the pipeline behavior, and the implemented optimizations gave the expected improvements. It took 1102.5h node hours in total. We processed 130TB of \textit{FASTQ} data and acquired an average mapping rate between 57\%-87\% depending on the tissue. 99\% of the input data have been computed with the pipeline without any errors. The 1\% accounts for errors such as out of memory for \textit{BAM} sorting step.

\begin{figure}[H]
    \centering
    \includegraphics[width=\linewidth]{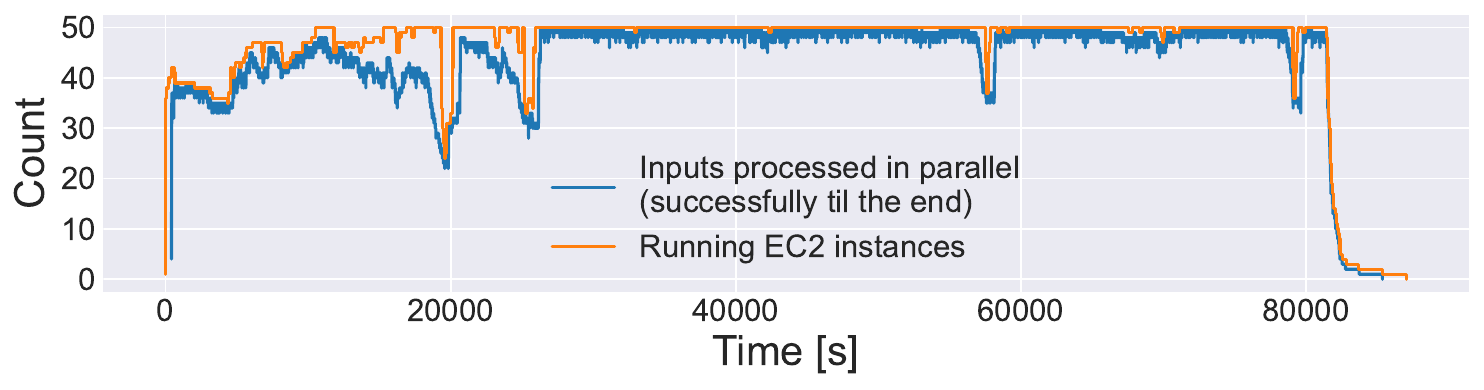}
    \caption{Large experiment timeline for instances and successfully computed files.}
    \label{fig:experiment_timeline}
\end{figure}

We estimate that the early stopping  reduces the total run time of STAR by about 23\%. Using spot instances saved 50\% the compute costs, but 138 interruptions occurred, wasting 2. 9\% of the total instances' run time. The lack of available spot instances can be seen at the start of the experiment in Fig.~\ref{fig:experiment_timeline}. 

\begin{figure}[H]
    \centering
    \begin{subfigure}[h]{0.45\linewidth}
        \centering
        \includegraphics[width=1\textwidth]{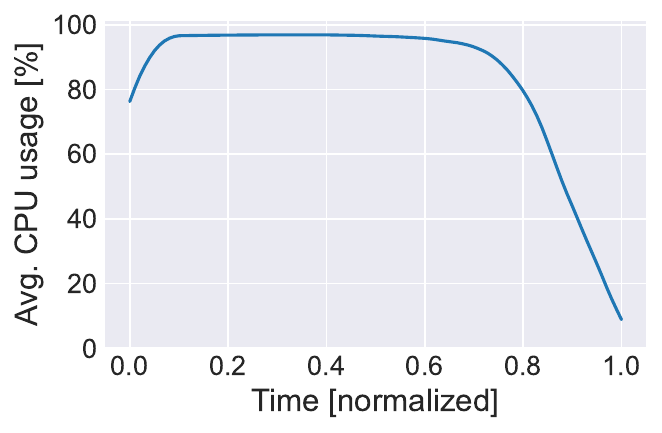}
    \end{subfigure}
    ~ 
    \begin{subfigure}[h]{0.45\linewidth}
        \centering
        \includegraphics[width=1\textwidth]{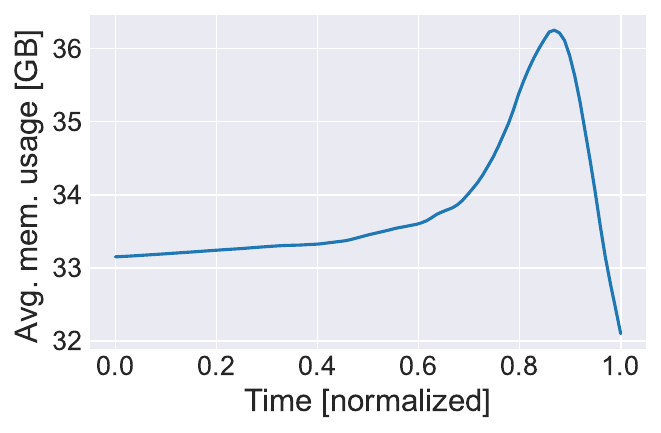}
    \end{subfigure}
    \caption{STAR aggregated and normalized metrics for CPU and memory usage.}
    \label{fig:STAR_footprint}
\end{figure}

The average CPU utilization across all instances was about 58\% for the entire pipeline and 78\% for STAR exclusively. 93\% of all instances run time the RAM utilization was between 45\% and 55\%, and only 1.7\% required more than 60\% of the instance memory suggesting area for improvement. STAR accounted for 71\% of the total workload time. In Fig.~\ref{fig:STAR_footprint}, we show aggregated CPU and memory usage during STAR for normalized metrics gathered during alignments longer than 10mins (n=1091). 

With implemented optimizations, the estimated cost is about \$477 - including compute (70\%), storage (18.5\%) and data transfer costs (11.5\%). This is equivalent to about \$0.066 per \textit{FASTQ} file. We estimate that an improved solution that would not require data transfer, along with optimizations of the size of the EBS, could reduce costs by an additional 20\%.

\vspace{-1.3cm}

\section{Conclusions}
\label{sec:conclusions}

This work presented cloud architecture for the Transcriptomics Atlas pipeline with the STAR aligner as its core. The optimizations described here significantly increased performance and throughput.
Analysis of a large-scale experiment showed that early stopping saved 23\% of the total running time of STAR. For use cases with a high mapping rate threshold, this feature would be even more beneficial. In addition, we observe that the pipeline is a great fit for spot instances and reduces computational costs by 50\%-60\%. 
Analysis of STAR's efficiency and resource utilization will help to choose the right configuration in order to maximize the throughput in similar scenarios. We identified one of the best instance types on the AWS cloud for running alignment in terms of processing time and cost.
The proposed solution proved to be cost-efficient and scalable, but we also identified areas for further improvements. 
Many of the insights in this work are applicable outside the cloud environment, extending the research results for HPC centers and workstations.
As concluded in~\cite{kica2024optimizing}, a faster STAR alignment can also improve the time required for a clinician to make a diagnosis.

\section{Future work}
\label{sec:future}

\subsection{Serverless containers as workers}
Using serverless services such as Google Cloud Run or Elastic Container Service in Fargate mode can be useful in certain scenarios. By adapting the current solution for that execution mode, we expect better scalability and resource allocation flexibility. However, using serverless usually comes with higher cloud costs, which can be acceptable if the goal is to process the input dataset as fast as possible. This approach is also beneficial to reduce operational costs and maintenance after the input dataset has been processed, but we may want to extend it with an additional dataset, which would be a magnitude order smaller.

\subsection{Double queue architecture}
Currently, the pipeline has a single queue for large and small tasks. By selecting input files that are smaller than arbitrary threshold (e.g. \textit{SRA} file $<$ 10GiB) we can reduce disk size and performance requirements. In addition, larger tasks are more affected by spot interruptions. Therefore, adding another queue for large tasks can improve resource utilization and reduce costs. This approach also reduces RAM usage for tasks that do not require a lot of memory during \textit{BAM} sorting if there is a way to estimate it beforehand.

\pagebreak

\subsection{Improvements of STAR index distribution }
Using NFS for the index distribution proved to be a viable solution. However, we have identified an underutilization of resources in the post-initialization phase. Using managed solutions such as Elastic File System or S3-express one-zone could speed up workers' start-time and reduce cross-AZ data transfer costs. S3 Express is a new storage class built for performance and is claimed to support hundreds of thousands of requests per second, which could improve the current solution.

\begin{credits}
\subsubsection{\ackname} The publication is supported by the Polish Minister of Science and Higher Education, contract number MEiN/2023/DIR/3796;  EU Horizon 2020 Teaming grant agreement No 857533; IRAP program of the Foundation for Polish Science MAB PLUS 2019/13; and EU Horizon Europe grant NEARDATA No 101092644.
\end{credits}

%

\bibliographystyle{splncs04}
\bibliography{references}

\end{document}